# *TMEM240* mutations cause spinocerebellar ataxia 21 with mental retardation and severe cognitive impairment

**Authors:**

Jérôme Delplanque, [1,2] David Devos, [3] Vincent Huin, [2,4] Alexandre Genet, [4] Olivier Sand, [1] Caroline Moreau, [3] Cyril Goizet, [5] Perrine Charles, [6] Mathieu Anheim, [7] Marie Lorraine Monin, [8] Luc Buée, [2] Alain Destée, [3] Guillaume Grolez, [3] Christine Delmaire, [9] Kathy Dujardin, [3] Delphine Dellacherie, [10] Alexis Brice, [11,12] Giovanni Stevanin, [11,13] Isabelle Strubi-Vuillaume, [4] Alexandra Durr, [11,12] Bernard Sablonnière, [2,4]

Correspondence to: Bernard Sablonnière, INSERM U837, Jean-Pierre Aubert Research Center, 1 Place de Verdun, F-59045, Lille, France. E-mail : bernard.sablonniere@inserm.fr

1 Centre National de la Recherche Scientifique, UMR8199, Institut de Biologie, F-59000, Lille, France

2 Institut National de la Santé et de la Recherche Médicale, INSERM U837, and Université de Lille Nord de France, F-59045, Lille, France

3 Département de Neurologie, Centre Hospitalier Régional et Universitaire de Lille, F-59037, Lille, France

4 Pôle de Biochimie et Biologie moléculaire, Centre de Biologie-Pathologie, Centre Hospitalier Régional et Universitaire de Lille, F-59037, Lille, France

5 Laboratoire des maladies rares : Génétique et métabolisme (MRGM/EA4576), Université Bordeaux Segalen, F-33076, Bordeaux, France

6 Département de Neurologie, Hôpital de la Pitié-Salpétrière, Assistance Publique-Hôpitaux de Paris, F-75651, Paris, France

7 Institut de Génétique et de Biologie moléculaire et cellulaire (IGBMC), INSERMU964/CNRS-UMR7104/ Strasbourg, F-67404, Illkirch-Graffenstaden, France 8 Département de Génétique et de Cytogénétique, Hôpital de la Pitié-Salpêtrière, Assistance Publique-Hôpitaux de Paris, F-75013, Paris, France

9 Département de Neuroradiologie, Centre Hospitalier Régional et Universitaire de Lille, F-59037, Lille, France

10 Département de Neuropédiatrie, Centre Hospitalier Régional et Universitaire de Lille, F-59037, Lille, France

11 Fédération de Génétique, Hôpital de la Pitié-Salpêtrière, Assistance Publique-Hôpitaux de Paris, F-75013, Paris, France

12 Institut du cerveau et de la moelle épinière, INSERM U1127/CNRS UMR7225, Université de Paris VI, F-75013, Paris, France

13 Laboratoire de Neurogénétique, Ecole Pratique des Hautes Etudes, Hésam Université, F-75013 Paris, France.

# ABSTRACT

Autosomal dominant cerebellar ataxia corresponds to a clinically and genetically heterogeneous group of neurodegenerative disorders that primarily affect the cerebellum. Here, we report the identification of the causative gene in spinocerebellar ataxia 21, an autosomal-dominant disorder previously mapped to chromosome 7p21.3-p15.1. This ataxia was firstly characterized in a large French family with slowly progressive cerebellar ataxia, accompanied by severe cognitive impairment and mental retardation in two young children. Following the recruitment of 12 additional young family members, linkage analysis enabled us to definitively map the disease locus to chromosome 1p36.33-p36.32. The causative mutation, (c.509C4T/p.P170L) in the transmembrane protein gene *TMEM240*, was identified by whole exome sequencing and then was confirmed by Sanger sequencing and co-segregation analyses. Index cases from 368 French families with autosomal-dominant cerebellar ataxia were also screened for mutations. In seven cases, we identified a range of missense mutations (c.509C4T/p.P170L, c.239C4T/p.T80M, c.346C4T/p.R116C, c.445G4A/p.E149K, c.511C4T/p.R171W), and a stop mutation (c.489C4G/p.Y163*) in the same gene. TMEM240 is a small, strongly conserved transmembrane protein of unknown function present in cerebellum and brain. Spinocerebellar ataxia 21 may be a particular early-onset disease associated with severe cognitive impairment.

**Key words:** Spinocerebellar ataxia; *TMEM240*; exome

**Abbreviations:** SCA = spinocerebellar ataxia; SPATAX = network of hereditary forms of spastic paraplegias and cerebellar ataxias

# INTRODUCTION

Autosomal-dominant spinocerebellar ataxia (SCA) corresponds to a clinically and genetically heterogeneous group of inherited neurodegenerative diseases that are characterized by progressive cerebellar ataxia. In some cases SCA is associated with pyramidal and extrapyramidal symptoms, cognitive impairment and peripheral neuropathy (Klockgether, 2011). To date, 37 different genetic loci have been reported as being associated with SCA subtypes, and 22 causative genes have been identified (Durr, 2010; Hersheson *et al.*, 2012; Matilla-Duenas *et al.*, 2012). In previous work, we characterized a 24-member French family pedigree, 11 members of which were affected by a new form of SCA (Devos *et al.*, 2001). The disease locus was initially mapped to chromosome 7 and designated as SCA21 (Vuillaume *et al.*, 2002). After examination of 12 additional family members, we reconsidered our linkage analysis. Here, we report on our use of whole exome capture to both localize and identify the causative gene. We identified a coding mutation in the *TMEM240* gene in the original SCA21 pedigree. Screening of 368 French families from the SPATAX network (http://spatax.wordpress.com) with autosomal-dominant cerebellar ataxia also detected this mutation in two other unrelated families. A variety of other mutations were also identified in members of five other families in the SPATAX cohort.

# MATERIALS AND METHODS

**Subjects**

We enrolled 37 members of Family A (the original SCA21 family), including 16 affected individuals. The age at onset ranged from 1 year to 21 years. Clinical severity was evaluated on the validated 40-point Scale for the Assessment and Rating of Ataxia (with 0 being normal) (Schmitz-Hubsch *et al.*, 2006). A cohort of 368 additional unrelated SCA pedigrees from the SPATAX cohort (originating mainly from France) was further analysed. In all families studied, a diagnosis of SCA was determined following a neurological examination by two or more experienced neurologists. Repeat expansions in the SCA1, 2, 3, 6, 7, 8, 12 and 17 genes were in the normal range and the presence of conventional mutations in SCA13, SCA14 and SCA28 genes was ruled out. Written informed consent was obtained from all subjects and the present study protocol was approved by the local investigational review board (registration number: DGS 2000/0468; approval for the SPATAX cohort: RBM 01-29).

Genomic DNA was isolated from peripheral blood mononuclear cells according to standard procedures.

**Genetic analysis**

*Linkage analysis*

In Family A, linkage analysis was performed on 27 family members using Illumina 16K infinity SNPlex microarrays (Illumina). Genotypes were determined using the genotyping module of Bead Studio software (Illumina) and analysed with MERLIN 1.0. software. Linkage analysis was performed under a 0.85 penetrance frequency for family members aged 11 years or older and 0.70 for those aged 10 years or under. We used an autosomal-dominant model and a disease frequency of 0.0001. To further narrow the candidate region, linkage analysis was refined via six microsatellite markers in the telomeric region of chromosome 1: a polymorphic dinucleotide repeat (CA22) located at 1.58 Mb, a polymorphic dinucleotide repeat (TG21) located at 2.05 Mb, D1S243 at 2.1 Mb, D1S468 at 3.58 Mb, D1S2845 at 4.3Mb and D1S2870 at 6.2 Mb. Lastly, a multipoint analysis was performed using the FastLink v4.1 software and GeneHunter v2.1 platform.

*Exome sequencing*

Whole exome sequencing was performed using an Agilent SureSelectXT Human Exon capture kit (Agilent Technologies). The enriched samples were sequenced on the HiSeq 2000 platform (Illumina). Three affected members and one unaffected member of Family A (Patients II-12, II-13, II-15 and III-12) were sequenced. Sequence reads were mapped to the reference human genome (UCSC NCBI37/hg19) using ELAND v2.0 software. Variant detection was performed with CASAVA v1.8 software and filtered to fit a minimum depth of 8. Variants were then annotated with Ensembl database (version 66) using the latter's Perl Application Program Interface: only non-synonymous coding, stop gain or loss, frame-shift, splice site and miRNA variants were kept for further analysis. The selected variants found in all three affected members but not in the unaffected member were annotated with dbSNP v135 and dbNSFP 2.0 beta1 (in silico functional predictions and allele frequencies from the 1000 Genomes Project).

**Molecular analyses of *TMEM240***

Mutation analysis of the four coding exons and flanking introns in *TMEM240* was performed with PCR followed by direct Sanger sequencing on an ABI3730XL sequencer (Applied Biosystems). Missense variations were analysed with Alamut 2.0 software, and protein multiple sequence alignments with Crustal Omega (Sievers *et al.,* 2011). The modified protein was also analysed according to the Uniprot database (http://www.uniprot.org/), the predicted protein structure was analysed using JPred, Tmpred, GOR4, Phyre2, and NetTurnP softwares. Phosphorylation sites were predicted using Phospho.ELM BLAST.

# RESULTS

**SCA21 is linked to the telomeric region of chromosome 1 in 1p36.33-p36.32**

Inclusion of additional family members and extensive clinical examination allowed us to reconstruct the family pedigree. Following genome-wide low density single nucleotide polymorphism analysis, only one chromosomal region reached the expected value of 3.0. This region spans ≈8Mb (starting from chromosome 1p) and gives a two-point logarithm (base 10) of odds (LOD) score of 3.35. Using six additional markers for fine mapping, significant linkage of the disease was found with four microsatellite markers (the CA22 repeat, TG21 repeat, D1S243 and D1S468, with LOD scores of 6.64, 2.77, 6.64 and 4.94, respectively. A maximum two-point LOD score (Zmax) of 6.64 was obtained for CA22 repeat and for D1S243 at a recombination fraction of 0.00. Three markers (CA22, TG21 and D1S243) enabled us to construct a unique haplotype shared by all affected individuals but none of the unaffected individuals. The critical interval was situated between the telomere end and the centromeric marker D1S468. In Subject III-21, the interval was defined by a recombination event distal to marker D1S243 (delimiting a 3.58Mb region). According to the GRCh37/hg19 genome reference, this locus includes 94 reference genes.

**Exome sequencing identifies the causative gene for SCA21**

We sequenced the exomes of four affected individuals from Family A. Using the de novo assembly of exon sequences, we detected 226 coding variants and 205 indels. After analysis and filtering, 11 heterozygous variants (six coding variants and five other variants) that mapped to 1p36.33-p36.32 were found to be shared by all affected subjects and were not present in control samples or in public databases. Sanger sequencing of these variants in the remaining 32 family members identified only one variant (c.509C4T/p.P170L in *TMEM24*0) that completely segregated with the disease phenotype (Fig. 1). This mutation was not found in 934 French controls subjects or in public databases.

**Six different mutations in the *TMEM24*0 gene**

To further confirm that the identified mutation was responsible for the SCA21 subtype, we sequenced the four coding exons of *TMEM240* gene in probands of 368 additional French unrelated autosomal-dominant SCA pedigrees from the SPATAX cohort for whom known SCA gene mutations had been ruled out. Six different mutations were found in seven additional unrelated families (Fig. 1). The same c.509C4T/ p.P170L in exon 4 was found in Families B and C. A c.489C4T/p.Y163* stop mutation in exon 4 was identified in Family D. Other missense mutations were as follows: c.346C4T/p.R116C in exon 3 in Family E, c.239C4T/p.T80M in exon 3 in Family F, c.511C4T/p.R171W in exon 4 in Family G and c.445G4A/p.E149K in exon 4 in Family H. Surprisingly, the mutations found in Families B and E seemed to be de novo events. A single founder mutation responsible for P170L in Families A, B and C seemed to be unlikely, since affected patients carrying the mutation did not bear the same haplotype (data not shown). In Families A, B and E, all subjects were available for DNA genotyping enabling us to confirm the complete segregation of the mutation with the disease. In Families D, F and G, the genotyping results (when available), showed that mutations were present in all affected subjects but in none of the unaffected relatives. In Families C and H, blood samples from siblings or children were not available for segregation analysis. Furthermore, none of the six mutations were found to be present in 396 unaffected controls of French ancestry and in thousands of control samples in public databases: [dbSNP, the 1000 Genomes Project, and the Exome Variant Server (http://evs.gs.washington. edu/EVS/)].

**Clinical features of patients with *TMEM240* mutations**

Since the first report (Devos *et al.,* 2001), 13 years of medical follow-up and the examination of additional children in Family A have enabled us to specify the most striking features of the disease (Table 1). Firstly, disease onset occurs early with the delayed acquisition of cognitive and motor skills. Secondly, the most striking sign is mild to severe mental retardation early evidenced at school with a variable degree of frontal behaviour disorders (impulsivity, aggressive, apathy, etc.). Thirdly, the predominant motor sign is clumsiness, which appears at a variable age (from age 2 to 20). Fourthly, the disease progresses very slowly. There was no motor impairment, cranial nerve palsy, pyramidal sign, fasciculation or sensory loss. There was no evidence of episodic cerebellar dysfunction or epilepsy in any affected family members. Neuropsychological examinations of patients revealed a severe cognitive disorder, including moderate impairments in attention, executive function, short-term, working and episodic memory abilities and, marked impairments in action planning, abstract reasoning, language and visuospatial functions. Given that most patients had marked reading and writing difficulties, the assessment procedure was adapted to include appropriate tests (Benton *et al.,* 1978; De Renzi and Faglioni, 1978; Wechsler *et al.,* 2004; Korkman *et al.,* 2007) (Table 2). Motor, visual and brainstem auditory evoked potentials and the results of an EMG and EEG examination were normal. In Family B, Patient III-1 presented with delayed acquisition of gait (after the age of 2 years), mild mental retardation and a severe cognitive impairment. In Family C, proband II-2, displayed moderately severe gait ataxia, writing difficulties and a mild intellectual disability. In Family D, two brothers (Patients III-2 and III-3) developed mild to moderate gait ataxia, in childhood. In Family E, proband II-2 presented delayed acquisition of cognitive and motor skills, moderate mental retardation and severe cognitive impairment. In Families F and G, the age at onset was late (59 and 61 years, respectively) and gait ataxia was mild to moderate. The proband in Family G, Patient II-2 had mild cognitive impairment. In Family H, proband III-1 displayed poor balance, dysarthria and gait difficulties at the age of 40. Cerebellar ataxia was usually quite severe and was associated with early or late onset cognitive impairment. MRI (3 T) of the brain revealed a cerebellar atrophy of the vermis and (to a lesser extent) the hemispheres in all families; although the brainstem was unaffected. Mild iron overload (as measured by $R2^*$ proton relaxation rates in MRI), was noted in several cases, and appeared to be most severe in the red nucleus and the pallidum in Families B and C. In Family G, brain MRI also revealed T2 and FLAIR hyperintensities in the pons, the pallidum, the putamen and the periventricular white matter.

**SCA21-associated mutations alter highly conserved amino acid residues in *TMEM240***

Alignment of orthologous proteins (from the zebrafish to human) revealed that the amino acids residues affected by the SCA-associated mutations were strongly conserved (Fig. 2A). Five of the six mutations concerned the carboxyl terminal portion of TMEM240. Three of these (Y163*, P170L and R171W) are clustered near a predicted beta turn centred on Proline

168 and Serine 169; the latter is phosphorylated in mouse brain (Huttlin, *et al.,* 2010; Trinidad *et al.,* 2012). Only T80M is located between the two transmembrane alpha helices (close to Tyrosine 77, which is predicted to be phosphorylated in human brain) (Fig.2B).

## DISCUSSION

We have identified a novel causative gene for SCA, *TMEM240*, with various missense mutations and a stop mutation in eight different autosomal-dominant SCA families. All the amino acids involved have been strongly conserved during evolution (from the zebrafish to the human). The original mutation (P170L, observed in three different families) was not observed in the Exome Variant Server or 1868 chromosomes from French controls. The other mutations were also unknown and not observed in 792 chromosomes from French controls. The T80M mutation (observed in Family F) is associated with a milder phenotype (i.e. the absence of cognitive impairment). These findings strongly suggest that *TMEM240* is the causative gene in SCA21. *TMEM240* is highly expressed in the brain in general and, in the cerebellum, dentate gyrus, putamen and caudate nucleus in particular (Hawrylycz *et al.,* 2012). *TMEM240* encodes a predicted transmembrane protein of unknown function but which has been detected in the mouse brain synapse membrane (Trinidad *et al.,* 2012). Another transmembrane protein (TMEM237) is involved in Joubert syndrome-related disorders characterized by a specific midhind brain malformation with hypoplasia of the cerebellar vermis (Huang *et al.,* 2011). TMEM240 does not have any homology with TMEM237 or any known protein or motif. We did not find any mutation that would undoubtedly have resulted in the loss of gene function. On the contrary, the only identified nonsense mutation that may have resulted in a gain-of-function with production of a truncated protein is Y163*. Indeed, this mutation, located in the last exon, might not be able to activate the nonsense-mediated decay pathway. Furthermore, the fact that all other are missense mutations supports the gain-of-function hypothesis in SCA21. Although mutations in *TMEM240* were found to be fully penetrant in six families, de novo mutations occurred in Families B and E; this could indicate the presence of spontaneous events in this telomeric region of chromosome 1. The *TMEM240* gene spans 5.6 kb on 1p36.33 and contains only four coding exons. It is lost in all pure terminal deletions and most other chromosomal rearrangements observed in 1p36 deletion syndrome (Heilstedt *et al.,* 2003; Gajecka *et al.,* 2007). In autosomal dominant cerebellar ataxia, mild mental retardation was recently described in SCA13 and mild to moderate late-onset cognitive impairment has been observed in SCA14 (Herman-Bert *et al.,* 2000; Stevanin

*et al.,* 2004). In contrast to SCA13, no petit mal epilepsy has been observed in SCA21. Subcortical dementia is, however, present in a small subgroup (10%) of patients with SCA1, SCA2, and SCA7 (Rüb *et al.,* 2013). Mild cognitive impairment and behavioural problems were observed in a SCA27 pedigree (Brusse *et al.,* 2006). In SCA29 (overlapping with SCA15), a similar phenotype was observed: congenital onset, no progression of ataxia and moderate cognitive impairment in both children and adults (Dudding *et al.,* 2004). In conclusion, mutations in a novel gene (*TMEM240*) cause SCA21, a rare form of mild cerebellar ataxia. These mutations had a frequency of 2% (8 of 368) in the French SPATAX cohort of cerebellar ataxia families analysed in this report. The most consistent feature of the disease was mild to-severe cognitive impairment (observed in seven of the eight families). Moreover, the associated severe cognitive impairment (with mild to moderate mental retardation in early life) is unusual; it's very slow progression sets it apart from other ataxias. We therefore recommend testing for SCA21 in families with mild or slowly progressive ataxia, particularly when moderate to severe cognitive impairment is present.

## ACKNOWLEDGEMENTS

We are indebted to all the patients and family members for their participation in this study. We are grateful to Emmanuelle Durand, Patrick Devos, Christiane Marzys, Kathy Dupont, Thomas Bardyn and Edwige Vanbrussel for excellent technical assistance; Elisabeh Vangelder and Alix de Becdelievre for their invaluable help in fine mapping; Elodie Petit and the biological resource centre at the Institut du Cerveau et de la Moelle Epinie` re for DNA from the SPATAX network. We thank Dr. David Fraser (Biotech Communication SARL, Damery, France) for helpful comments on the manuscript's English.

## FUNDING

This work was funded by the University of Lille 2, the Institut National de la Santé et de la Recherche Médicale (INSERM), the European Union (6[th] Framework Programme for Research and Technological Development: the EUROSCA consortium), and the Centre Hospitalier Régional et Universitaire de Lille (CHRU). The linkage and exome work was funded by the Genome Analysis Facility at the Lille Biology Institute, and University of Lille 2. Fine genotyping analysis and Sanger sequencing was performed by the Lille Biology and Pathology Centre (Molecular Biology Core Facility). Part of this work was funded by the 7[th] framework program on OMICS (the NEUROMICS consortium), the French government's "Investissements d'avenir" program (reference ANR-10-IAIHU-06, awarded to Institut du Cerveau et de la Moelle Epinière, and the Fondation Roger de Spoelberch.

# TABLES

| Family code | A | | | | | | | B | | C | D | E | F | G | H |
|---|---|---|---|---|---|---|---|---|---|---|---|---|---|---|---|
| Subject code | II-15 | II-18 | III-13 | III-14 | III-15 | III-20 | III-26 | II-3 | III-1 | II-2 | III-2 | II-2 | II-2 | II-2 | III-1 |
| Sex | F | M | M | M | F | M | M | F | F | F | M | M | M | F | F |
| Mild delayed acquisition | NA | NA | + | + | + | + | + | + | + | + | + | + | NA | NA | NA |
| Age of recognised ataxia, y | 10 | 18 | 1 | 1 | 1 | 9 | 2 | 20 | 2 | 14 | NA | 21 | 59 | 61 | 40 |
| Age of $1^{st}$ examination, y | 34 | 32 | 1 | 1 | 1 | 17 | 2 | 56 | 19 | 38 | 25 | 25 | 63 | 63 | 50 |
| Gait/limb ataxia | ++/++ | +/+ | ++/+ | ++/+ | ++/+ | ++/+ | ++/+ | ++/++ | +/+ | +/+ | +/+ | +/+ | +/+ | +/+ | ++/++ |
| Tremor (rest/postural) | +/+ | -/- | -/- | -/- | -/- | +/+ | +/+ | +/+ | +/+ | -/+ | +/+ | +/+ | -/- | -/- | -/- |
| Tendon Reflexes | ↘ | ↗ | ↘ | ↘ | ↘ | ↗ | ↗ | ↘ | ↘ | N | N | ↗ | N | ↗ | ↗ |
| Babinski sign | - | + |  | - | - | + | + | - | - | + | - | + | - | - | + |
| Dysarthria/dysgraphia | +/+ | +/+ | +/+ | +/+ | +/+ | +/+ | +/+ | +/+ | +/+ | +/+ | +/+ | +/+ | +/- | -/- | +/- |
| Ocular findings | Ny,S | Ny,S | Ny,S | Ny,S | Ny,S | Ny,S | Ny,S | Ny,S | Ny,S | D | Ny,S | Ny,S,D | S | Ny,D | Ny |
| SARA score | 14 | 8 | 10 | 11 | 9 | 9 | 7 | 26 | 14 | 10 | 10 | 13 | 17 | x | 20 |
| Brain MRI | CA | CA | N | N | N | CA | N | CA | CA | CA | CA | CA | CA | CA | CA |
| Behavior troubles | + | ++ | + | + | + | ++ | + | + | ++ | ++ | + | + | - | + | - |
| IQ/ID | 50 | NA/++ | 54 | 46 | 50 | 64 | 60 | NA/+ | NA/++ | NA/+ | NA/+ | 51 | NA/- | NA/ | NA/+ |

**Table 1.** Clinical features in 15 affected members from the eight families

N = normal; NA = not assessed. Clinical signs are graded as follows: _ = absent; + = Moderate; + + = severe. Ny = nystagmus; S = slow ocular saccades; D = diplopia; CA = cerebellar atrophy of the vermis and (to a lesser extent) the hemispheres. ID = intellectual disability (noted in the absence of a measured IQ).

|  | II-13 | II-15 | III-12 | III-13 | III-14 | III-15 |
| --- | --- | --- | --- | --- | --- | --- |
| Age at examination, y | 54 | 47 | 25 | 13 | 12 | 9 |
| Abstract reasoning[a] | + | +++ | ++ | + | + | +++ |
| Visual working memory[b] | + | +++ | ++ | N | +++ | +++ |
| Visual episodic memory[c] | +++ | +++ | +++ | N | ++ | + |
| Oral comprehension[d] | ++ | ++ | ++ | + | + | +++ |
| Executive functions[e,g] | + | +++ | ++ | + | + | +++ |
| Visuospatial functions[f] | +++ | +++ | ++ | ++ | +++ | +++ |
| Speed of information processing[g] | ++ | +++ | ++ | +++ | +++ | +++ |
| Selective attention[g] | ++ | +++ | +++ | +++ | +++ | +++ |

**Table 2.** Cognitive impairment test results for six patients from Family A

N = normal; + = mild; + + = moderate; + + + = severe impairment.

[a] Assessed by performance in the Coloured Progressive Matrices and the Wechsler Intelligence Scale for Children IV's Picture Concepts for children.

[b] The Corsi block-tapping test.

[c] The 10/36 test in adults and the Children's Memory Scale Dot Location test in children.

[d] TheToken Test in adults and the NEPSY-II Comprehension of Instructions subtest in children.

[e] The Tower of London Test.

[f] Benton Judgement of Line Orientation in adults and the NEPSY-II Arrows subtest in children.

[g] A computer-controlled task

**FIGURE LEGENDS**

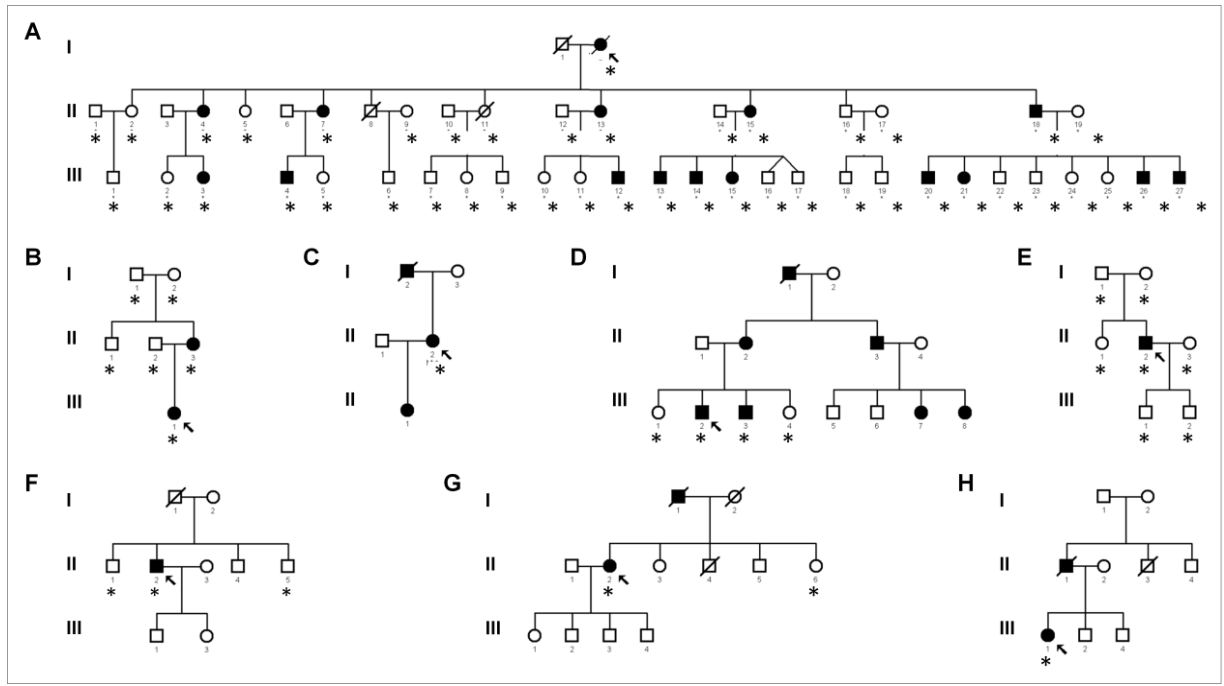

**Figure 1.** Pedigree charts of the eight SCA21 families.

The eight families correspond to the following codes: (A) LIL-001; (B) LIL-002; (C) AAD-617; (D) AAD-881; (E) LIL-003; (F) AAD-1005; (G) AAD-656; and (H) AAD-725. Circles denote females; squares denote males. The probands are denoted by an arrow. Filled symbols denote clinically affected members, open symbols indicate unaffected individuals. / = deceased. *Genotyped individuals.

.

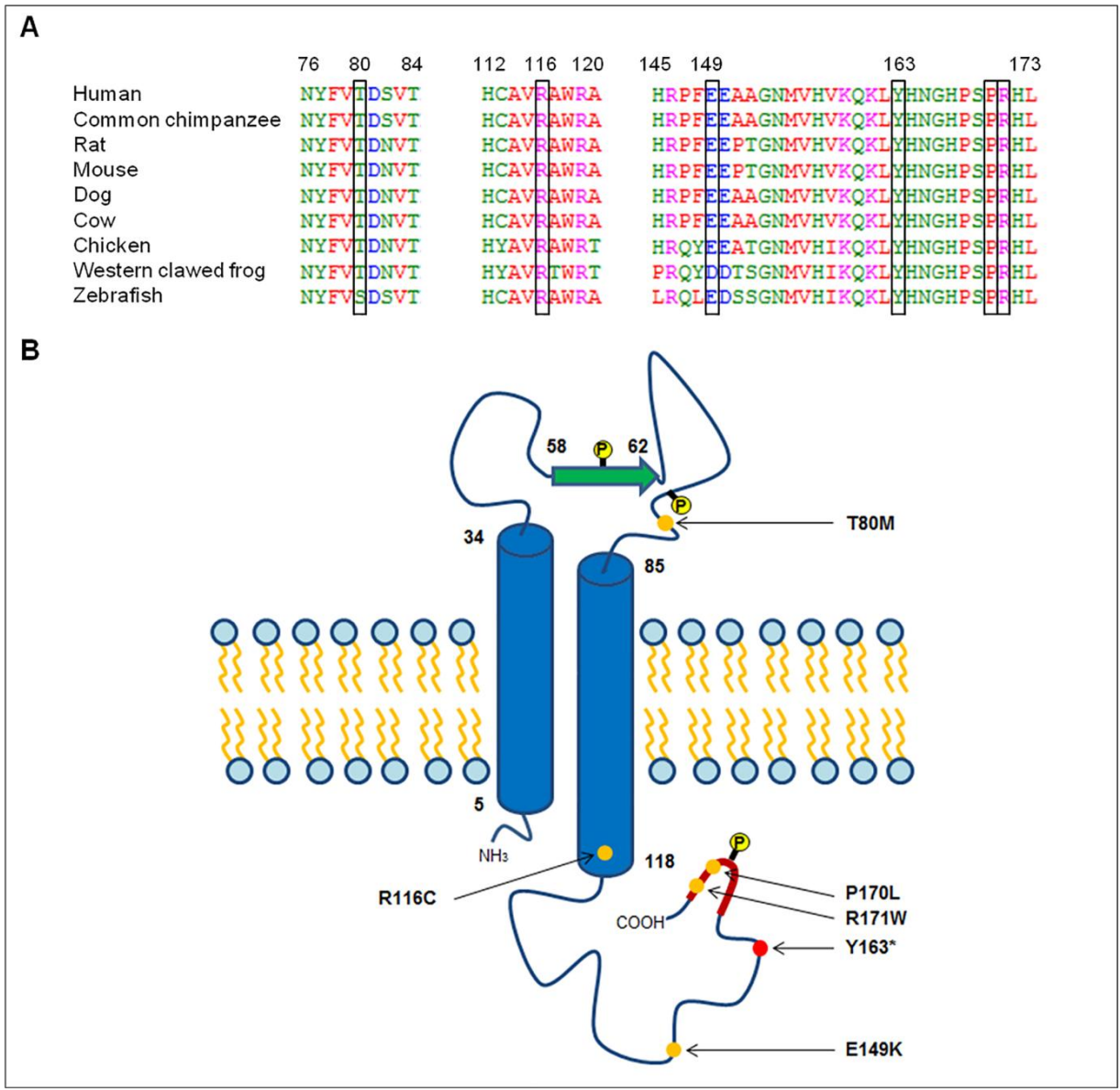

**Figure 2.** Nature and position of six mutations within the predicted protein TMEM240.
(A) The mutated residues have been conserved throughout evolution, as shown by alignment of the protein sequences of TMEM240/C1orf70 orthologues in various organisms with Crustal Omega software. (B) TMEM240's predicted secondary structure, showing the location of the six mutations. Predicted alpha-helices are depicted as cylinders, beta-sheets as green arrows and beta turns as red bold lines. Probable transmembrane helices are filled in purple. Yellow dots indicate the predicted phosphorylated residues Y60, Y77 and S169. Orange dots indicate missense mutations and the red dot indicates a nonsense mutation.

# SUPLEMENTARY DATA

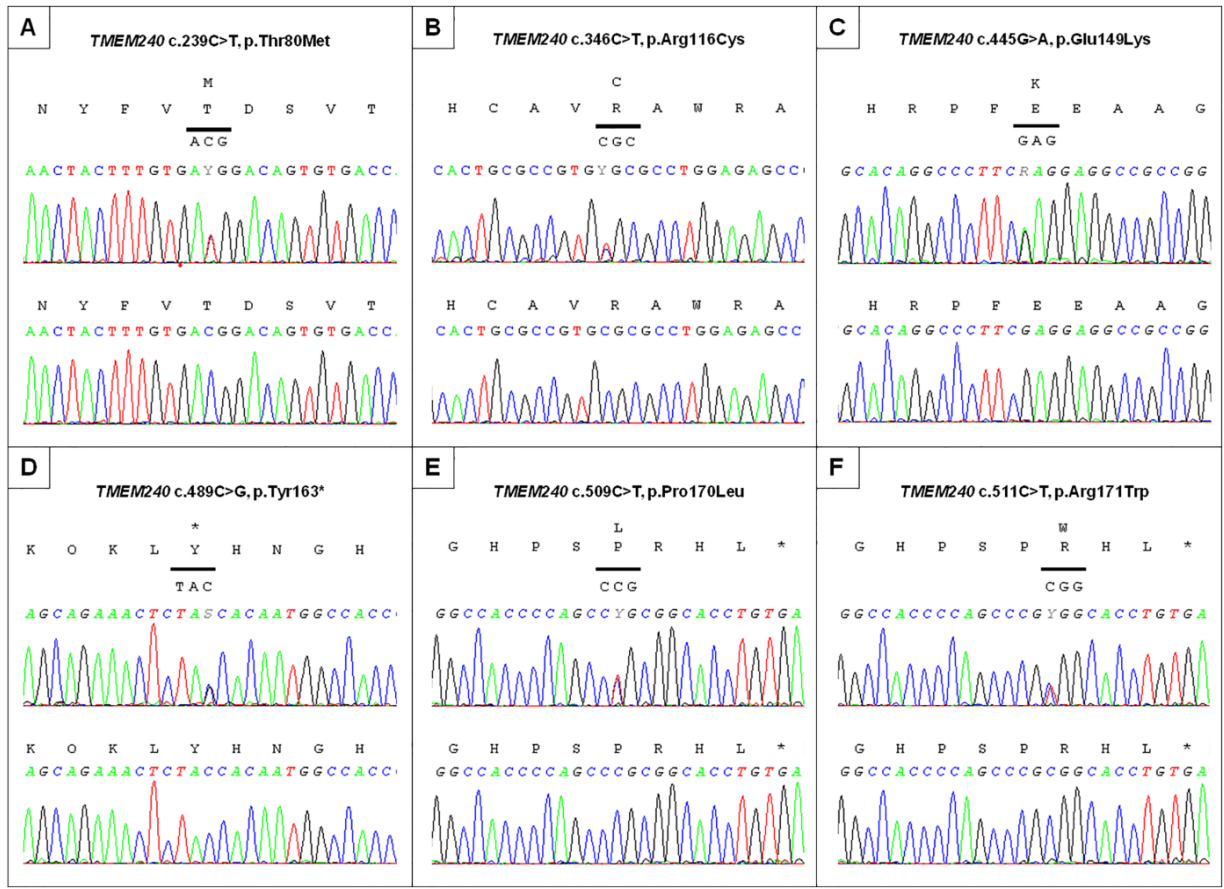

**Supplemental Figure 1.** Electropherograms of *TMEM240* heterozygous mutations.

Electropherograms of a patient (up) and a control (down) **A.** T80M mutation (sporadic case, family F). **B.** R116C (sporadic case, family E). **C.** E149K (proband only, family G). **D.** Y163* (family D). **E.** P170L (families A, B and C). **F.** R171W (proband only, family H).